\documentstyle[preprint,aps]{revtex}
\begin{document}
\draft

\title{Adiabatic Theory of Electron Detachment from Negative Ions in 
Two-Color Laser Field}

\author{M.~Yu.~Kuchiev and V.~N.~Ostrovsky \cite{SP}}

\address{School of Physics, University of New South Wales,
Sydney 2052, Australia}

\maketitle

\begin{abstract}

Negative ion detachment in bichromatic laser field is considered 
within the adiabatic theory. The latter represents a recent 
modification of the famous Keldysh model for multiphoton ionization 
[L.~V.~Keldysh, Zh. Eksp. Teor. Fiz. {\bf 47}, 1945 (1964)
[Sov. Phys.-JETP {\bf 20}, 1307 (1965)]]
which makes it quantitatively reliable. We calculate
angular differential detachment rates, partial rates
for particular ATD (Above Threshold Detachment) channels
and total detachment rates for H$^-$ ion in a bichromatic field
with $1:3$ frequency ratio and various phase differences.
% The effects of phase difference between the fields are reproduced. 
Reliability of the present, extremely simple approach is 
testified by comparison with much more elaborate earlier calculations. 

\end{abstract}


%\twocolumn
%\narrowtext

\section{Introduction} \label{Sec1}

Interest to the photoionization of atoms in a bichromatic laser 
field both in theory
(see, for instance, Ref. \cite{Barexp}--\cite{Cionga})
%\cite{Bartheo} \cite{Sz} \cite{SK} \cite{Pot91} \cite{Pot92} 
%\cite{Pot94} \cite{Bar93}\cite{Paz} \cite{Pro} \cite{T} \cite{Maq} 
and in experiment 
\cite{Barexp} \cite{Muller}--\cite{Y}
%\cite{Ce} 
seems to stem first of all from the effect of the phase control, i.e.
dependence of the observables on the difference of field 
phases $\varphi$. 

The calculations have been carried out previously for {\it ionization}\/ 
of the hydrogen atom in two laser fields with a  frequency ratio 
$1:2$ \cite{SK}, $1:3$ \cite{Pot91} and $2:3$ \cite{Pot94}.
Potvliege and Smith \cite{Pot92} presented results for various 
frequency ratios and initial states. Different schemes have 
been employed, but all of them imply numerically intensive work.

For the multiphoton electron {\it detachment}\/ from negative 
ions some analytical treatment exists \cite{Bar93} \cite{Paz} 
which aims to investigate qualitative features of the process,
mostly in the case when one or both fields are weak. 
The presence of large number of parameters in the problem
sometimes makes results of analytical studies not directly
transparent. Quantitative reliability of these approaches 
has never been assessed. This situation looks particularly 
unsatisfactory since the multiphoton electron detachment from 
negative ions presents unique situation when quantitative results 
can be obtained by analytical method in a broad range of 
parameters characteristic to the problem. Indeed, it has been 
demonstrated recently by Gribakin and Kuchiev \cite{GKa} \cite{GKb} 
that proper application of the well-known Keldysh \cite{Keldysh}
model to multiphoton detachment \cite{ftnt} provides 
an extremely simple scheme that gives very reliable results
for the {\it total rates}\/ as well as for ATD (Above Threshold 
Detachment) {\it spectrum}\/ and ATD {\it angular distributions}.
This {\it adiabatic approximation}\/ ensures an accuracy which is
comparable with that of the most elaborate numerical developments
and works unexpectedly well even outside its formal applicability
range, i.e. even {\it for small number $n$ of photons absorbed}. 
The evidences of good performance of the Keldysh model for 
the {\it total rates}\/ were presented also in the 
earlier paper \cite{Shake}.

Recently the adiabatic approach was extended by the present 
authors \cite{KO} to the case of bichromatic field. The practical
applications were carried out for the case of frequency ratio
$1:2$ when in addition to the phase effects another unusual
phenomenon exists, namely the {\it polar asymmetry}\/ of the angular 
distribution of photoelectrons. Unfortunately no other
quantitative data for photodetachment in this case is available
which makes comparison impossible.

The main objective of the present study is to assess 
quantitatively an accuracy of the adiabatic scheme by 
comparison with the previous calculations carried out by 
Telnov {\it et al}\/ \cite{T} in case of $1:3$ frequency ratio.
For this ratio the polar asymmetry is absent, but the phase
effects persist. The calculations by Telnov {\it et al}\/ \cite{T}
are based on sufficiently sophisticated numerical scheme providing
a useful benchmark. We present (Sec. \ref{Sec3}) 
complete comparison of the results by considering angular 
differential detachment rates, heights of ATD 
peaks and total detachment rates. It should be emphasized 
that the angular differential rates are most sensitive 
to the formulation of the model representing an ultimate 
test for the theory, as discussed in Sec.\ref{Sec4}.
We draw also some general conclusion on the 
relation between the adiabatic approach and the numerical 
calculations within the one-electron approximation.

\section{Results} \label{Sec3}

The adiabatic theory of two-color detachment was outlined 
in our previous paper \cite{KO} where the reader can find
all the details of calculation. Here we only write down
the expression for the electric field strength
in the bichromatic laser field with $1:3$ frequency ratio
in order to specify the definition of the field phase difference
$\varphi$
\begin{eqnarray} \label{F}
\vec{F}(t) = \vec{F}_1 \, \cos \omega t +
\vec{F}_2 \, \cos ( 3 \omega t + \varphi ) ~.
\end{eqnarray}
$\vec{F}_1$, $\vec{F}_2$ are the amplitude vectors for 
the fundamental frequency $\omega$ and its third harmonics 
respectively. Below we consider, just as in Ref.\cite{T}, the case 
when both fundamental field and its third harmonics are linear 
polarized with $\vec{F}_1 \parallel \vec{F}_2$. Then the 
differential photoionization rate depends only on the single 
angle $\theta$ between the photoelectron translational momentum 
$\vec{p}$ and the vectors $\vec{F}_1 \parallel \vec{F}_2$. 
Atomic units are used throughout the paper unless stated otherwise.

Our calculations for H$^-$ detachment are carried out for the 
parameters of H$^-$ as before \cite{KO} ($\kappa = 0.2354$, 
$A=0.75$).
We choose two sets of field intensities
$I_1$ and $I_2$ for the fundamental frequency $\omega = 0.0043$ 
(that of CO$_2$ laser) and its third harmonics, 
same as in the paper by Telnov {\it et al }\/ \cite{T}, namely 
(i) $I_1 = 10^{10} {\rm W}/{\rm cm}^2$, $I_2 = 10^{9} {\rm W}/{\rm cm}^2$
and 
(ii) $I_1 = 10^{10} {\rm W}/{\rm cm}^2$, $I_2 = 10^{8} {\rm W}/{\rm cm}^2$.

In case of the frequencies ratio $1 : 3$
considered here the field (\ref{F}) does not possess polar 
asymmetry (i.e. asymmetry under inversion of the $z$ axis
directed along $\vec{F}_1 \parallel \vec{F}_2$).
Therefore the differential detachment rate 
does not change under the transformation 
$\theta \Rightarrow \pi - \theta$. This allows us to
show plots only for $\frac{1}{2} \pi \geq \theta \geq 0 $
domain.

Figs. \ref{Fig1}-\ref{Fig6} show the differential detachment 
rate as a function of the angle $\theta$ for three 
lowest (open) ATD channels and for two sets of field intensities. 
The system Hamiltonian is a 2$\pi$-periodic function of the phase
parameter $\varphi$. We show our results for $\varphi = 0$,
$\varphi = \pm \frac{1}{2}$ and $\varphi = \pi$. 
The transformation $\varphi \Rightarrow \pi - \varphi$ leaves 
the Hamiltonian invariant only if $t$ is replaced by $-t$.
As stressed in Refs.\cite{Bartheo}, the problem is invariant
under the time inversion operation provided the final-state
electron interaction with the atomic core is neglected.
This is the case in the present model.
Therefore our differential ionization rates are the same for
$\varphi$ and $- \varphi$.
The calculations by Telnov {\it et al}\/ \cite{T} do take 
into account the final state electron-core interaction.
Therefore they show some difference between the angular 
differential rates for $\varphi$ and $-\varphi$. However, it
proves to be quite small for low ATD channels as seen 
from the plots. 

The importance of the interaction between the emitted
electron and the core has been first pointed out by one of the
present authors \cite{Ku87}. In this paper several phenomena 
has been predicted for which this interaction plays crucial role.
The related mechanism was named {\it ``atomic antenna''}.
In the recent literature the final state interaction is usually
referred to as {\it rescattering}. In our problem the
rescattering effects are enhanced for high ATD channels 
as discussed below. 

The results of our extremely simple theory are compared in 
Figs. \ref{Fig1}-\ref{Fig6} 
with the previous numerical calculations by 
Telnov {\it et al}\/ \cite {T} which are rather 
involving. Being carried out in the one-electron approximation,
they employ an accurate model for the effective one-electron 
potential in H$^-$ \cite{LC}, complex-scaling generalized pseudospectral 
technique \cite{CSGPS} to discretize and facilitate the solution
of the time-independent non-Hermitian Floquet Hamiltonian for
complex quasienergies and eigenfunctions, and calculation of
the electron energy and angular distributions by the reverse
complex-scaling method \cite{reverse}. As a lucid illustration
of simplicity of the present approach it is worthwhile to
stress that it does not rely on any particular form of
an effective one-electron potential albeit employs only two 
parameters $\kappa$ and $A$ governing the asymptotic behavior
of the initial bound state wave function. 

 From Figs. \ref{Fig1}-\ref{Fig6} one can see that 
the adiabatic approximation ensures
good quantitative agreement with calculations by Telnov 
{\it et al}\/ \cite{T}. In particular, positions of maxima and 
minima in the angular photoelectron
distribution are well reproduced. This demonstrates that the
adiabatic approach correctly describes the nature of the structure 
as due to interference between the electron waves emitted at 
various (complex-valued) moments of time. Indeed, within 
the adiabatic theory \cite{KO} the ionization amplitude
is expressed as a sum of a number of interfering contributions.
Mathematically they come from different saddle points in
the approximate evaluation of the integral over time that 
emerges in the Keldysh \cite{Keldysh} model. Physically they correspond 
to the coherent emission of photoelectron at different moments of time. 
For our particular frequency ratio $1:3$ the sum contains 6 interfering
contributions as compared with 4 term for $1:2$ frequency ratio 
\cite{KO} and 2 terms for one-color detachment \cite{GKa} \cite {GKb}.
Generally this suggests that in the former case more complicated
angular patterns emerge. Probably one can find here a correlation
with an alternative interpretation in the multiphoton
absorption framework. The latter argues \cite{Pot94} \cite{T}
that the angular distribution 
structure in $1:3$ case is more complicated than for $1:2$ ratio
since all the pathways leading to a continuum state with the 
same energy interfere in the $1:3$ case whereas a considerable pattern
of non-interfering pathways exists for the the $1:2$ case due to
parity or energy restrictions (each pathway is
characterized by the number of photons of different colors
absorbed successively).

The partial detachment rate for each ATD channel 
are shown in tables \ref{Tab1} and \ref{Tab2} for 
two sets of field intensities.
The agreement is good for low ATD channels; note that the 
rescattering effects which generate dependence on the sign of 
$\varphi$ are manifested in the partial rates even less than
in the angular distributions shown in Figs. \ref{Fig1}-\ref{Fig6}. 
For higher ATD channels with low rates the difference between 
the present results and those of Telnov {\it et al}\/ \cite{T} 
becomes more pronounced. This behavior could be interpreted as
increasing importance of rescattering for high ATD peaks. 
The manifestations of this effect were observed recently in experiment
\cite{e} and are currently vividly discussed in the literature
\cite{Ku87} \cite{e} \cite{t} \cite{Z}.

\section{Conclusion} \label{Sec4}

As a summary, the adiabatic approach provides quantitatively 
reliable tool for investigating two-color photodetachment 
of negative ions. In particular, the interference structure 
in the photoelectron angular distributions as well as
the phase effects are correctly described. Since 
generally the interference phenomena are known 
to be most sensitive to the details of theoretical
description, one can conclude that the present theory
had successfully passed the stringent test.

The Keldysh scheme \cite{Keldysh} is known to be gauge-noninvariant.
Importantly, the calculations within the adiabatic approach 
\cite{GKa} \cite{GKb} \cite{KO} employ the dipole-length 
gauge for the laser field thus stressing contribution of 
the long-range asymptote of the initial bound state wave function. 
The use of the length gauge together with the
adiabatic approach (i.e. integration over time by the saddle
point method, see Refs.\cite{GKa} \cite{GKb} \cite{KO}
and discussion in Sec.\ref{Sec3}) render self-consistent 
character to the theoretical scheme. Indeed, the exact 
evaluation of the integrals does not add to the accuracy 
of the result as compared with the use of the saddle point 
method. This is because in the former case the integral
absorbs the contributions from the wave function outside 
its asymptotic domain, where in fact it is known with much 
lower accuracy (being, in particular, influenced by the 
effects beyond the single active electron approximation).

The method is straightforwardly applicable to the negative 
ions with the outer electron having non-zero orbital momentum, 
such as halogen ions, which could be easier accessible 
for the experimental studies (for the one-color detachment 
such applications could be found in Ref.\cite{GKa}).
Technically the calculations within the adiabatic approach are
extremely simple reducing to finding the roots of polynomial
and substituting them into an analytical expression \cite{KO}
(the related {\it Mathematica} \cite{Math} program takes only 
few lines). It should be recognized that the single active 
electron approximation itself introduces some intrinsic error.
It seems that often this error could be comparable with the
difference between the result of numerical one-electron 
calculations and these of the adiabatic approximation. 
Uncertainty of the one-electron approach in principle could be 
removed within the two-electron approach which however
consumes much more efforts. The two-electron calculations
which has been carried out recently show that the one-electron 
approximation is generally sufficient unless one is particularly
interested in the subtle resonance effects \cite{Shake} 
\cite{Lamb} \cite{San}
(the calculations beyond one-electron approximation are
currently possible only for small number of absorbed photons).
For high ATD channels with low intensities the adiabatic
approximation becomes less reliable due to increasing role 
of rescattering effects neglected in the present form of
the approximation. It seems however that relatively simple
modifications of the adiabatic approximation could be carried 
to include rescattering effects. 

Reliability of the results obtained above for the simple one-electron 
problem with rescattering neglected is highly important in
perspective, since they are to be included as a constituent
part in the treatment of much more sophisticated one-electron 
and many-electron problems governed by the antenna mechanism
\cite{Ku87} \cite{Ku95} \cite{Ku96}.

\acknowledgements

We appreciate fruitful discussions with G.~F.~Gribakin. 
We are grateful to the referee of the present paper
for attracting our attention to Ref. \cite{Shake}.
The support from the Australian Research Council 
is thankfully acknowledged.
V.~N.~O. acknowledges a hospitality of the stuff of the
School of Physics of UNSW where this work has been
carried out.

\newpage

\begin{table} 
\caption{ \label{Tab1}
Partial rates for the H$^-$ detachment by the laser wave
with the frequency $\omega = 0.0043$ and its third harmonics with
the intensities  respectively $I_1 = 10^{10} {\rm W}/{\rm cm}^2$
and $I_2 = 10^{9} {\rm W}/{\rm cm}^2$.
The number of absorbed photons $n$ refers to the fundamental 
frequency. 
In each block the upper figure gives present result and
the lower one the result obtained by Telnov {\it et al }\/
\protect \cite{T}. 
The number in square brackets indicate the power of 10. 
}

\vspace{5mm}
\begin{tabular}{c c c c c c c}
    &             &           &           &           &          &  \\
    & One-colour  & \multicolumn{4}{c}{Two-colour} & One-colour \\
$n$ & fundamental & 
\multicolumn{4}{c}
{--------------------------------------------------------} 
& harmonic \\
    &           & $\varphi=0$ & $\varphi=\pi$ & $\varphi=\frac{1}{2}\pi$ &
$\varphi= - \frac{1}{2}\pi$ &  \\
    &            &           &           &           &           &  \\
\hline
    &            &           &           &           &           &  \\
8   & 0.67[--9]   & 0.47[--7]  & 0.54[--8]  & 0.22[--7]  & 0.22[--7]  &  \\
    & 0.72[--9]   & 0.42[--7]  & 0.58[--8]  & 0.20[--7]  & 0.21[--7]  &  \\
    &            &           &           &           &           &  \\
9   & 0.20[--9]   & 0.11[--7]  & 0.23[--8]  & 0.80[--8]  & 0.80[--8]  & 0.46[--10]\\
    & 0.20[--9]   & 0.10[--7]  & 0.23[--8]  & 0.71[--8]  & 0.73[--8]  & 0.30[--10]\\
    &            &           &           &           &           &  \\
10  & 0.41[--10]  & 0.27[--8]  & 0.34[--8]  & 0.39[--8]  & 0.39[--8]  &  \\
    & 0.39[--10]  & 0.26[--8]  & 0.27[--8]  & 0.35[--8]  & 0.30[--8]  &  \\
    &            &           &           &           &           &  \\
11  & 0.50[--11]  & 0.65[--9]  & 0.23[--8]  & 0.17[--8]  & 0.17[--8]  &  \\
    & 0.40[--11]  & 0.72[--9]  & 0.16[--8]  & 0.15[--8]  & 0.10[--8]  &  \\
    &            &           &           &           &           &  \\
12  & 0.74[--12]  & 0.16[--9]  & 0.10[--8]  & 0.62[--9]  & 0.62[--9]  & 0.66[--13]\\
    & 0.71[--12]  & 0.20[--9]  & 0.71[--9]  & 0.58[--9]  & 0.30[--9]  & 0.86[--13]\\
    &            &           &           &           &           &  \\
13  & 0.21[--12]  & 0.38[--10] & 0.36[--9]  & 0.20[--9]  & 0.20[--9]  &  \\
    & 0.33[--12]  & 0.53[--10] & 0.27[--9]  & 0.20[--9]  & 0.85[--10] &  \\
    &            &           &           &           &           &  \\
14  & 0.64[--13]  & 0.88[--11] & 0.11[--9]  & 0.58[--10] & 0.58[--10] &  \\
    & 0.14[--12]  & 0.14[--10] & 0.97[--10] & 0.69[--10] & 0.32[--10] &  \\
    &            &           &           &           &           &  \\
15  & 0.16[--13]  & 0.20[--11] & 0.32[--10] & 0.16[--10] & 0.16[--10] & 0.16[--15]\\
    & 0.47[--13]  & 0.32[--11] & 0.33[--10] & 0.23[--10] & 0.17[--10] & 0.31[--15]\\
    &            &           &           &           &           &  \\
16  & 0.36[--14]  & 0.52[--12] & 0.82[--11] & 0.44[--11] & 0.44[--11] &  \\
    & 0.14[--13]  & 0.72[--12] & 0.12[--10] & 0.78[--11] & 0.10[--10] &  \\
    &            &           &           &           &           &  \\
17  & 0.68[--15]  & 0.15[--12] & 0.21[--11] & 0.12[--11] & 0.12[--12] &  \\
    & 0.35[--14]  & 0.22[--12] & 0.43[--11] & 0.27[--11] & 0.60[--11] &  \\
    &            &           &           &           &           &  \\
Total & 0.92[--9] & 0.62[--7] & 0.15[--7] & 0.36[--7] & 0.36[--7] & 0.46[--7]\\
      & 0.96[--9] & 0.56[--7] & 0.14[--7] & 0.33[--7] & 0.33[--7] & 0.30[--7]\\

\end{tabular}
\end{table}

\begin{table}
\caption{ \label{Tab2}
Same as in table 1, but for the intensities 
$I_1 = 10^{10} {\rm W}/{\rm cm}^2$
and $I_2 = 10^{8} {\rm W}/{\rm cm}^2$.
}

\vspace{5mm}
\begin{tabular}{c c c c c c c}
    &             &           &           &           &          &  \\
    & One-colour  & \multicolumn{4}{c}{Two-colour} & One-colour \\
$n$ & fundamental & 
\multicolumn{4}{c}
{---------------------------------------------------------} 
& harmonic \\
    &           & $\varphi=0$ & $\varphi=\pi$ & $\varphi=\frac{1}{2}\pi$ &
$\varphi= - \frac{1}{2}\pi$ &  \\
    &            &           &           &           &           &  \\
\hline
    &            &           &           &           &           & \\
8   & 0.67[--9]   & 0.54[--8]  & 0.41[--9]  & 0.25[--8]  & 0.25[--8]  & \\
    & 0.72[--9]   & 0.53[--8]  & 0.36[--9]  & 0.24[--8]  & 0.25[--8]  & \\
    &            &           &           &           &           & \\
9   & 0.20[--9]   & 0.92[--9]  & 0.16[--9]  & 0.69[--9]  & 0.69[--9]  & 0.46[--13]\\
    & 0.20[--9]   & 0.93[--9]  & 0.18[--9]  & 0.68[--9]  & 0.68[--9]  & 0.30[--13]\\
    &            &           &           &           &           & \\
10  & 0.41[--10]  & 0.23[--9]  & 0.82[--10] & 0.25[--9]  & 0.25[--9]  & \\
    & 0.39[--10]  & 0.25[--9]  & 0.82[--10] & 0.25[--9]  & 0.22[--9]  & \\
    &            &           &           &           &           & \\
11  & 0.50[--11]  & 0.54[--10] & 0.74[--10] & 0.88[--10] & 0.88[--10] & \\
    & 0.40[--11]  & 0.59[--10] & 0.65[--10] & 0.88[--10] & 0.66[--10] & \\
    &            &           &           &           &           & \\
12  & 0.74[--12]  & 0.11[--10] & 0.38[--10] & 0.28[--10] & 0.28[--10] & 0.66[--17]\\
    & 0.71[--12]  & 0.14[--10] & 0.33[--10] & 0.28[--10] & 0.22[--10] & 0.98[--17]\\
    &            &           &           &           &           & \\
13  & 0.21[--12]  & 0.21[--11] & 0.13[--10] & 0.79[--11] & 0.79[--11] & \\
    & 0.33[--12]  & 0.45[--11] & 0.13[--10] & 0.82[--11] & 0.94[--11] & \\
    &            &           &           &           &           & \\
14  & 0.64[--13]  & 0.39[--12] & 0.39[--11] & 0.20[--11] & 0.20[--11] & \\
    & 0.14[--12]  & 0.21[--11] & 0.47[--11] & 0.24[--11] & 0.49[--11] & \\
    &            &           &           &           &           & \\
15  & 0.16[--13]  & 0.66[--13] & 0.96[--12] & 0.46[--12] & 0.46[--12] & 0.16[--21]\\
    & 0.47[--13]  & 0.11[--11] & 0.17[--11] & 0.70[--12] & 0.25[--11] & \\
    &            &           &           &           &           & \\
16  & 0.36[--14]  & 0.11[--13] & 0.21[--12] & 0.99[--13] & 0.99[--13] & \\
    & 0.14[--13]  & 0.52[--12] & 0.60[--12] & 0.21[--12] & 0.12[--11] & \\
    &            &           &           &           &           & \\
17  & 0.68[--15]  & 0.17[--14] & 0.42[--13] & 0.21[--13] & 0.21[--13] & \\
    & 0.35[--14]  & 0.22[--12] & 0.22[--12] & 0.64[--13] & 0.50[--12] & \\
    &            &           &           &           &           & \\
Total & 0.92[--9] & 0.66[--8] & 0.79[--9] & 0.36[--8] & 0.36[--8] & 0.46[--13]\\
      & 0.96[--9] & 0.66[--8] & 0.74[--9] & 0.35[--8] & 0.35[--8] & 0.30[--13]\\

\end{tabular}
\end{table}

%***************************************************

\begin{figure} 
\caption{ \label{Fig1}
Detachment of H$^-$ ion in bichromatic field
with the frequencies $\omega = 0.0043$ and $3\omega$
and intensities $I_1 = 10^{10} {\rm W}/{\rm cm}^2$
and $I_2 = 10^{9} {\rm W}/{\rm cm}^2$ respectively.
Differential detachment rate 
(in units $10^{-8}$a.u.) as a function of the electron
emission angle $\theta$
is shown for the first ATD peak (corresponding to absorption 
of $n=8$ photons of frequency $\omega$) and various values
of the field phase difference $\varphi$ as indicated in the plots.
Open symbols show results of calculations by Telnov {\it et al}\/
\protect\cite{T} 
(in the $\varphi = \pm \frac{1}{2} \pi$ plot the open  
circles show the results for $\varphi = \frac{1}{2} \pi$ and
open triangles these for $\varphi = - \frac{1}{2} \pi$).
Solid curves show results of the present adiabatic theory
(which coincide for $\varphi = \frac{1}{2} \pi$ and
$\varphi = - \frac{1}{2} \pi$ as discussed in the text).
}
\end{figure}

\begin{figure}
\caption{  \label{Fig2}
Same as in Fig.1, but for the second ATD peak (corresponding to
absorption of $n=9$ photons of frequency $\omega$).}
\end{figure}

\begin{figure}
\caption{  \label{Fig3}
Same as in Fig.1, but for the third ATD peak (corresponding to
absorption of $n=10$ photons of frequency $\omega$).}
\end{figure}

\begin{figure}
\caption{  \label{Fig4}
Same as in Fig.1, but for intensities $I_1 = 10^{10} {\rm W}/{\rm cm}^2$
and $I_2 = 10^{8} {\rm W}/{\rm cm}^2$;
the differential detachment rate is shown for the first ATD peak 
(corresponding to absorption of $n=8$ photons of frequency $\omega$).}
\end{figure}

\begin{figure}
\caption{  \label{Fig5}
Same as in Fig.4, but for the second ATD peak (corresponding to
absorption of $n=9$ photons of frequency $\omega$).}
\end{figure}

\begin{figure}
\caption{  \label{Fig6}
Same as in Fig.4, but for the third ATD peak (corresponding to
absorption of $n=10$ photons of frequency $\omega$).}
\end{figure}

\newpage
\begin{figure}[h]
\input psfig
\psfig{file=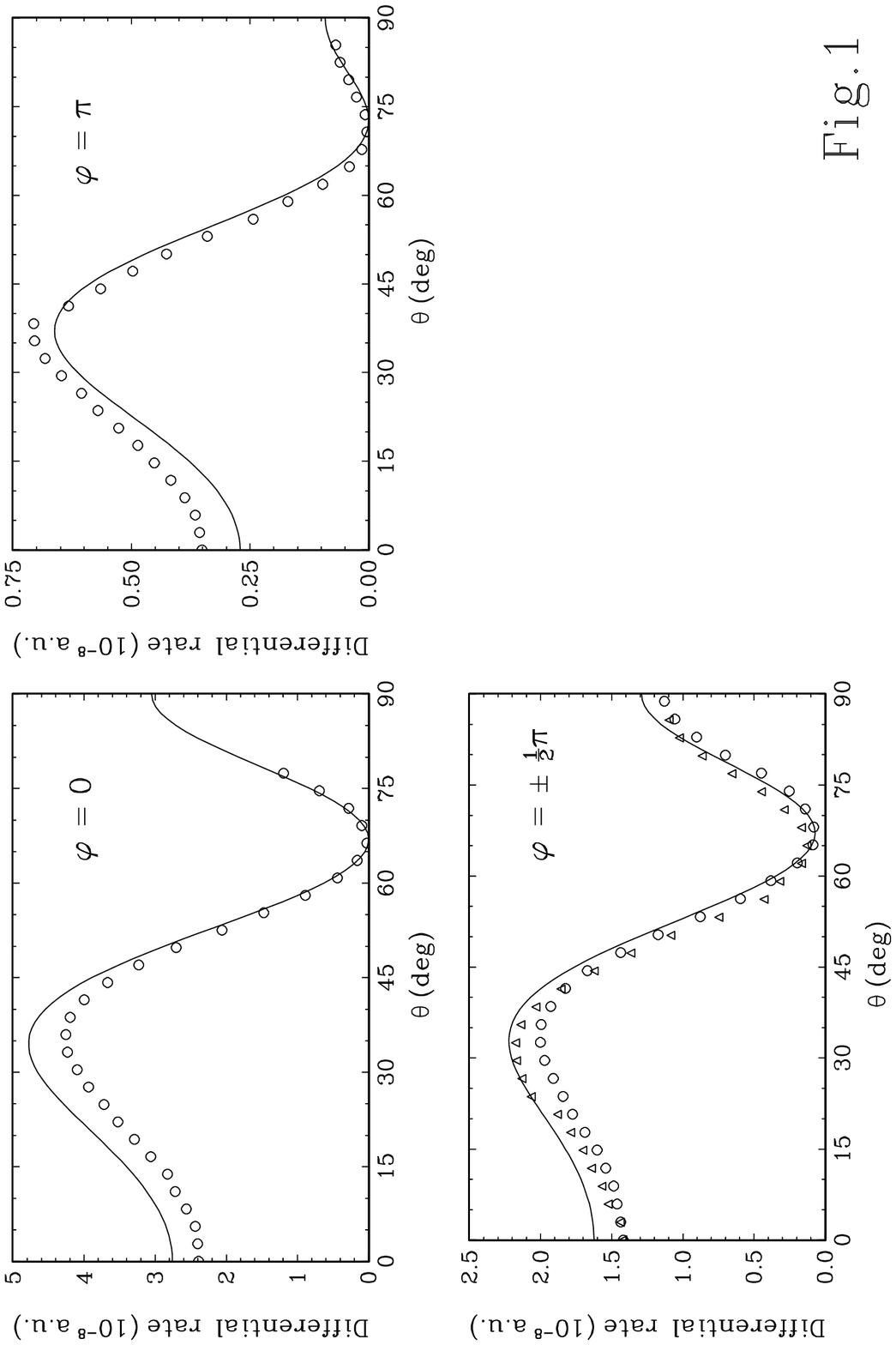, clip=}
\end{figure}
\newpage
\begin{figure}[h]
\input psfig
\psfig{file=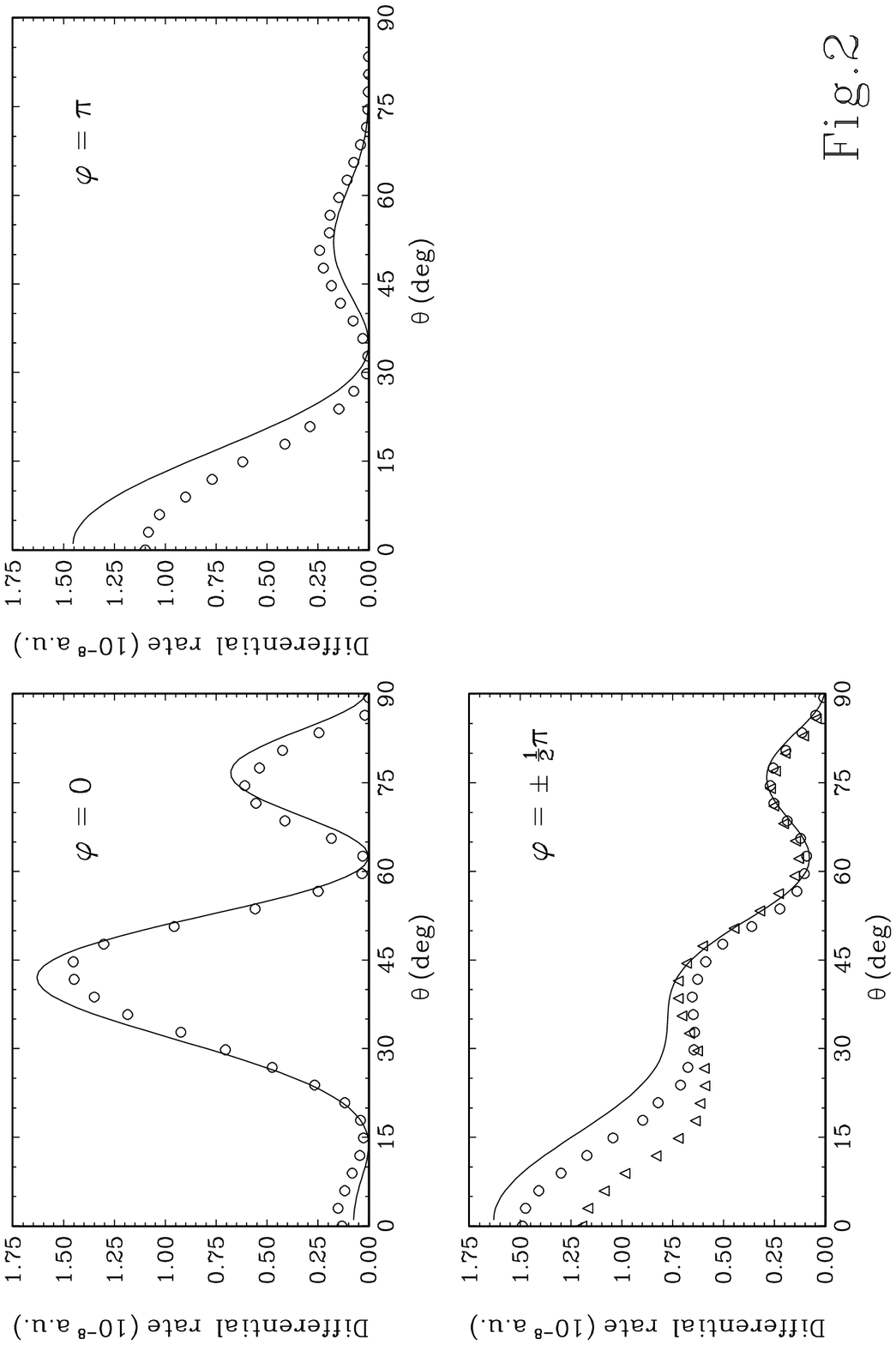, clip=}
\end{figure}
\newpage
\begin{figure}[h]
\input psfig
\psfig{file=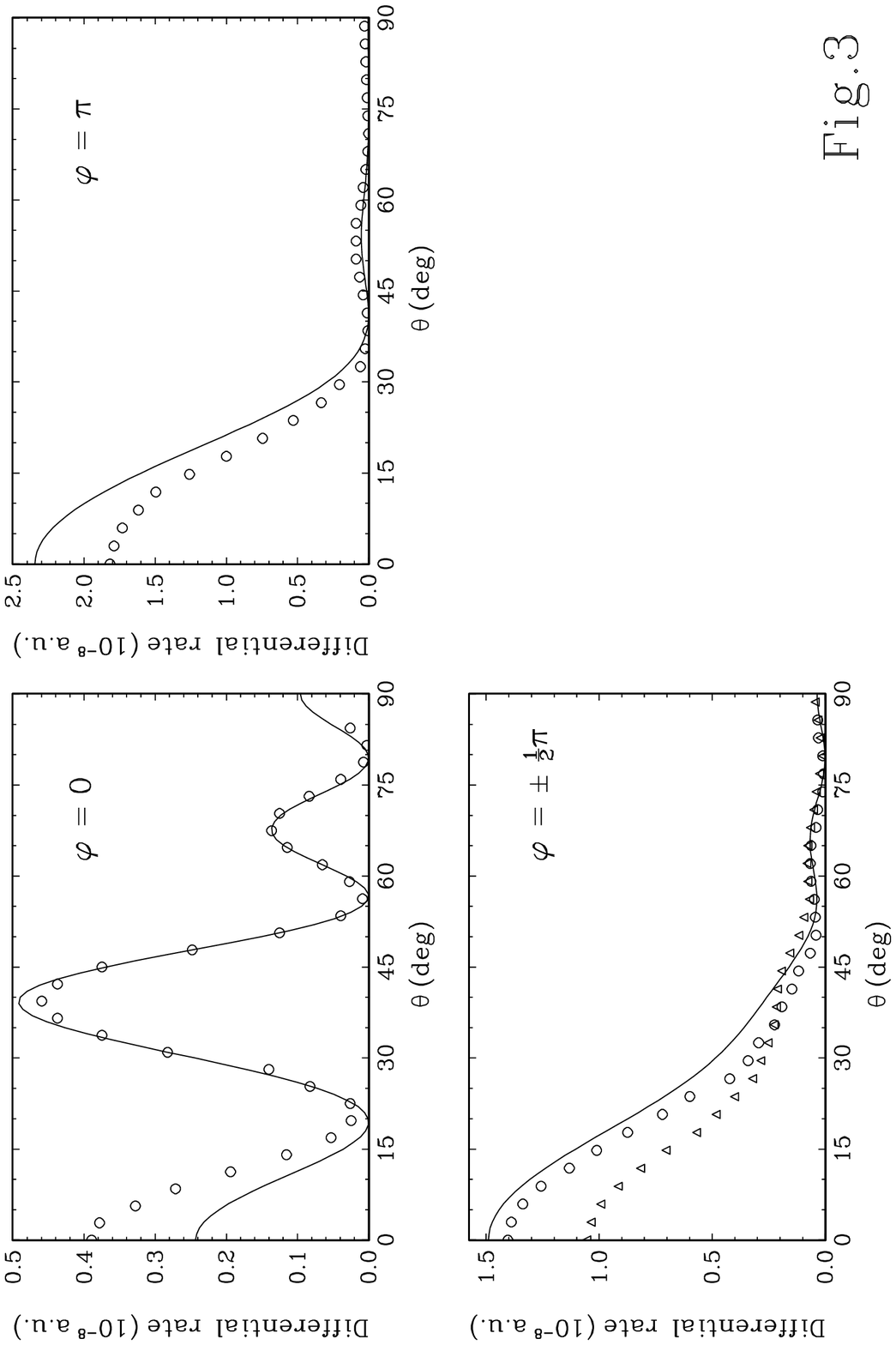, clip=}
\end{figure}
\newpage
\begin{figure}[h]
\input psfig
\psfig{file=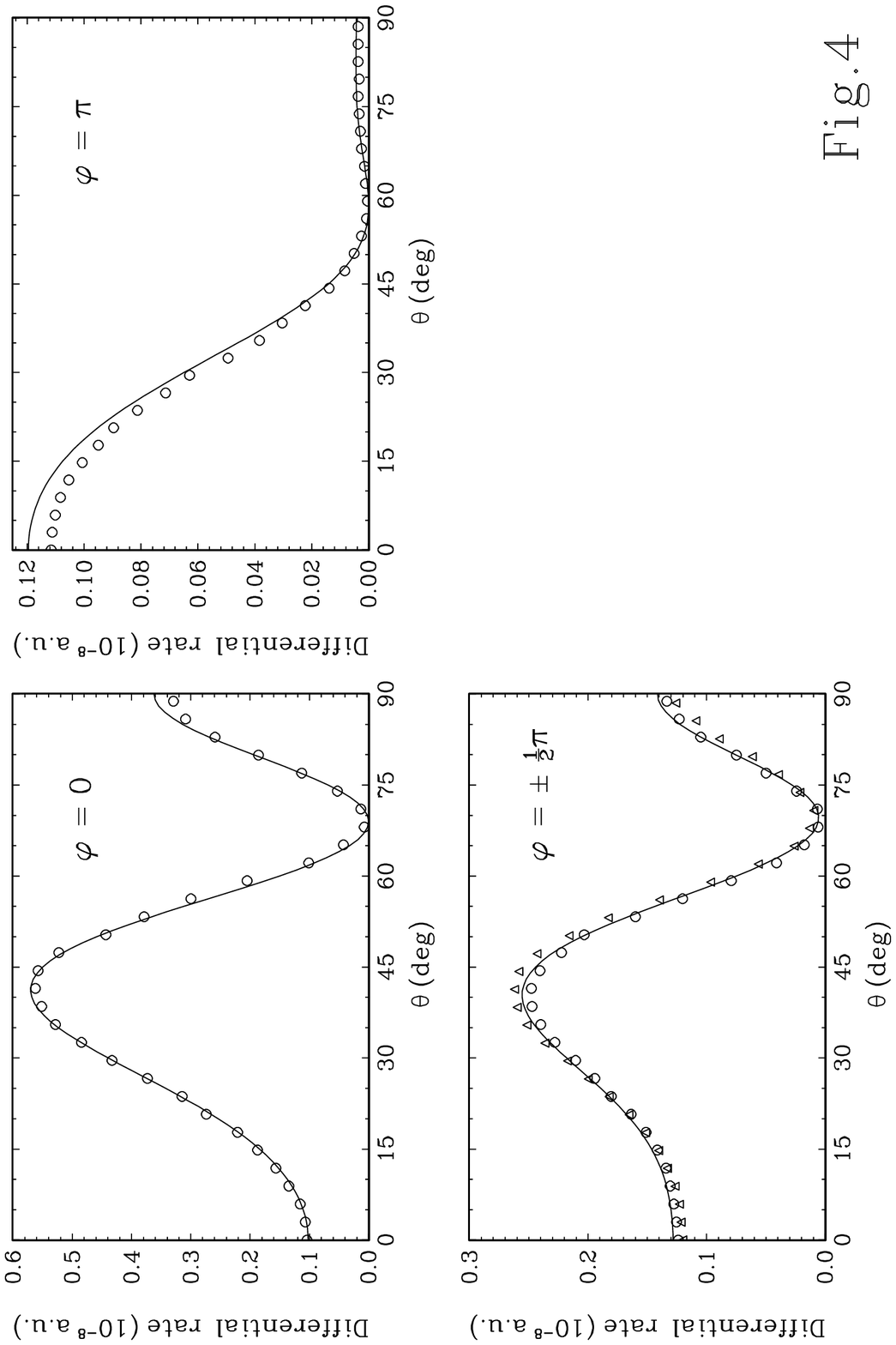, clip=}
\end{figure}
\newpage
\begin{figure}[h]
\input psfig
\psfig{file=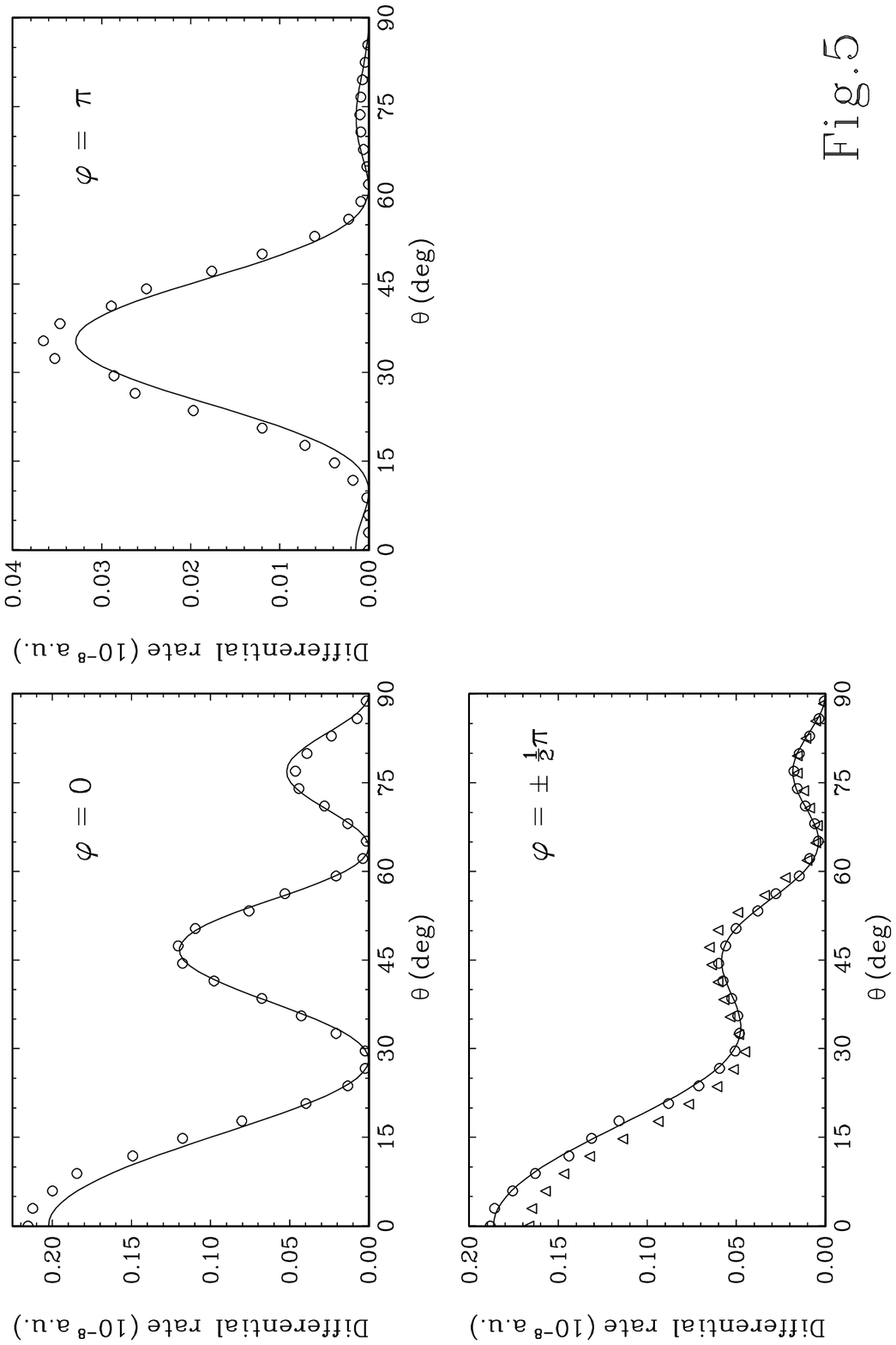, clip=}
\end{figure}

\newpage
\begin{figure}[h]
\input psfig
\psfig{file=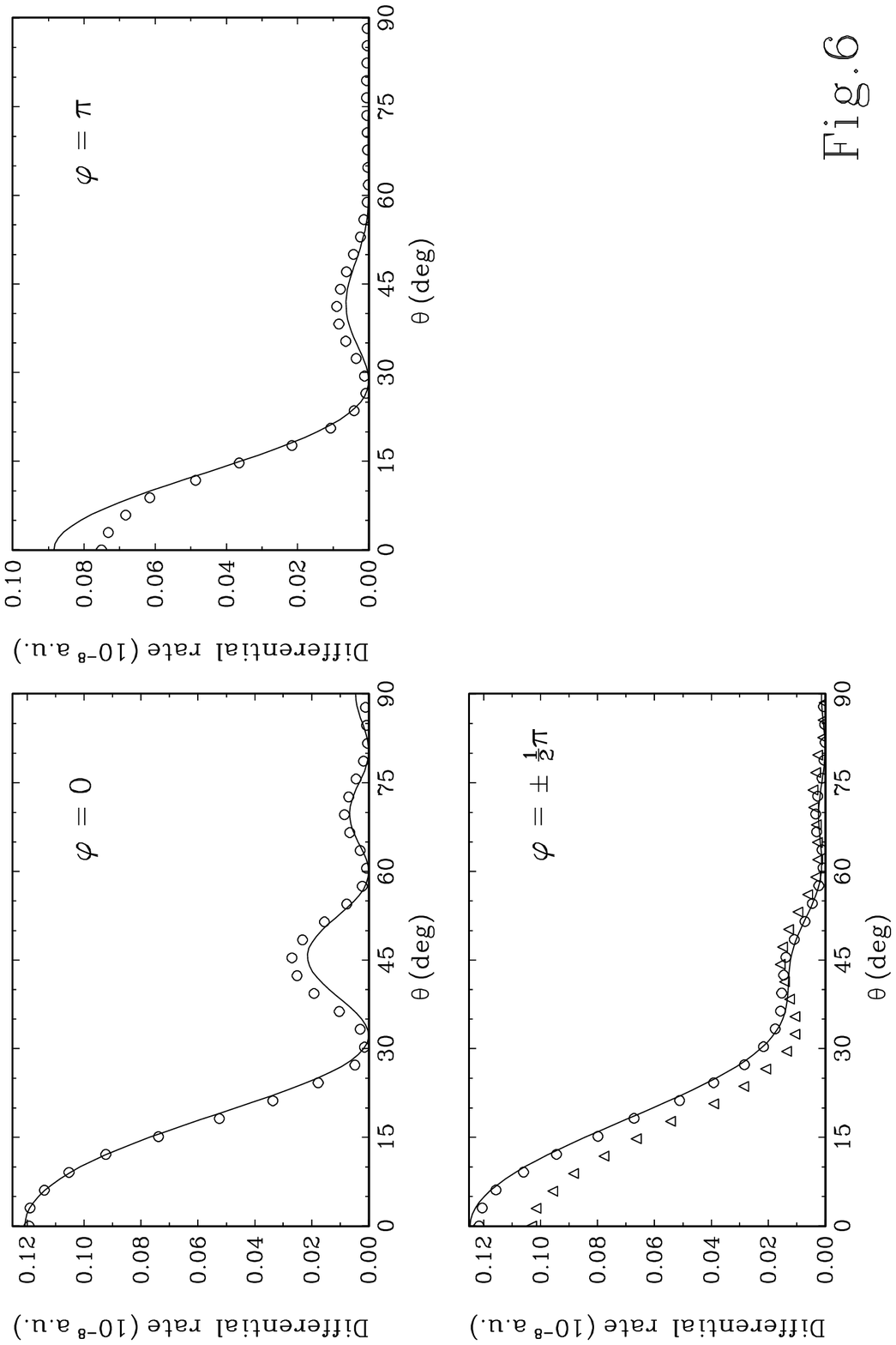, clip=}
\end{figure}

\end{document}